\documentclass[12pt]{article}
\usepackage{epsfig}

\newcommand{\bea}{\begin{eqnarray}}
\newcommand{\eea}{\end{eqnarray}}
\newcommand{\be}{\begin{equation}}
\newcommand{\ee}{\end{equation}}

\newcommand{\ar}{a_s}

\begin{document}

\begin{center}
  {\bf $\alpha_s$ in DIS scheme}

\vskip 0.5cm

A.V. Kotikov$^{a}$,
V.G.~Krivokhizhin$^a$, and B.G.~Shaikhatdenov$^a$ 

\vskip 0.5cm
${}^a$ Joint Institute for Nuclear Research, Russia
\end{center}


\begin{abstract}
Deep inelastic scattering data on $F_2$ structure function accumulated by various collaborations in
fixed-target experiments are analyzed in the nonsinglet approximation
  and within $\overline{MS}$
  and DIS schemes.
  The study of high statistics deep inelastic scattering data provided by BCDMS, SLAC and NMC
  collaborations, is carried out
  by applying a combined analysis.
  The application of the DIS scheme leads to the resummation of contributions that are important 
in the region of large $x$ values. 
  It is found that using the DIS scheme does not significantly change the strong coupling constant itself
 but does strongly change the values of the twist-four corrections.

\end{abstract}

$PACS:~~12.38~Aw,\,Bx,\,Qk$\\


\section{ Introduction }

Currently, the accuracy of data for the deep-inelastic scattering (DIS) structure functions (SFs) allows us to study
$Q^2$-dependence of logarithmic corrections based on QCD and
power-like (non-perturbative) corrections separately (see, for example, ~\cite{Beneke} and references therein).

Nowadays a commonly adopted benchmark tool for this kind of analyses turns out to be at the next-to-next-to-leading-order (NNLO) level
(see e.g.~\cite{PKK}-\cite{Kotikov:2015zda}  and references therein).
However, the relevant papers have recently been published in which QCD analysis of DIS SFs 
has been carried out up to the next-to-next-to-next-to-leading order (NNNLO)
\cite{Blumlein:2008kz,Iran}. 

This article is closely related to those devoted to similar studies, with the main difference being that here we work
within the so-called DIS scheme \cite{DIS}, whose application leads to effective resummation of large-$x$ logarithms 
into the Wilson coefficient functions.
We analyze DIS structure function $F_2(x,Q^2)$ with SLAC, NMC and BCDMS
experimental data~\cite{SLAC1}--\cite{BCDMS1} at NNLO level of massless perturbative QCD.

As in our previous papers \cite{Shaikhatdenov:2009xd,Kotikov:2015zda,KK2001}, the function $F_2(x,Q^2)$ is
represented as a sum of the leading twist $F_2^{\rm pQCD}(x,Q^2)$ and the twist-four terms
\footnote{Hereinafter, superscripts {\rm pQCD,~LT} denote the twist two approximation with and without 
target mass corrections (see, for example, \cite{KK2001}).}:
\be
F_2(x,Q^2)=F_2^{\rm pQCD}(x,Q^2)\left(1+\frac{\tilde h_4(x)}{Q^2}\right)\,.
\label{1.1}
\ee

\section{ Theoretical aspects of the analysis }

Here we briefly describe some aspects of the theoretical part of our analysis.
For a more detailed description, see ~\cite{KK2001,Shaikhatdenov:2009xd}.
In the large $x$-values region gluons do almost not contribute, and the $Q^2$ evolution of the
twist-two DIS  structure function $F_2(x,Q^2)$ is determined by the so-called nonsinglet (NS) part.

In this approximation, there
is a direct relation between the moments of the DIS  structure function $F_2(x,Q^2)$
and the moments of the NS parton distribution function (PDF) ${\bf f}(x,Q^2)$
\footnote{Unlike the standard case, here PDF is multiplied by $x$.}
\be
M_{n}(Q^2) ~=~\int_0^1 dx x^{n-2} F^{\rm LT}_{2}(x,Q^2),~~
{\bf f}(n,Q^2) ~=~\int_0^1 dx x^{n-2}\, {\bf f}(x,Q^2) 
\label{Moments}
\ee
in the following form
\be
M_n(Q^2) = R_{\rm NS}(f)\times C(n,\ar(Q^2))\times {\bf f}(n,Q^2)\,,
\label{3.a}
\ee
where the strong coupling constant
\be
\ar(Q^2)=\frac{\alpha_s(Q^2)}{4\pi} \label{as}
\ee
and $C(n,\ar(Q^2))$ stands for the Wilson coefficient function.
The constant $R_{NS}(f)$ depends on weak and electromagnetic charges and is fixed to one sixth for $f=4$~\cite{Buras}.

\subsection{Strong coupling constant}

The strong coupling constant is determined from the corresponding renormalization group equation.
At the NLO level, the latter is as follows
\bea \label{1.coA}
\frac{1}{a_{1}(Q^2)} - \frac{1}{a_{1}(M_Z^2)} +
b_1 \ln{\left[\frac{a_{1}(Q^2)}{a_{1}(M_Z^2)}
\frac{(1 + b_1a_{1}(M_Z^2))}
{(1 + b_1a_{1}(Q^2))}\right]}
= \beta_0 \ln{\left(\frac{Q^2}{M_Z^2}\right)}\,,
\eea
where
\be
a_{1}(Q^2)=\ar^{\rm NLO}(Q^2),~~a_{2}(Q^2)=\ar^{\rm NNLO}(Q^2) \, .
\label{ai}
\ee

At NNLO level, the strong coupling constant is derived from the following equation:
\bea \label{1.co}
\frac{1}{a_2(Q^2)} - \frac{1}{a_2(M_Z^2)} &+&
b_1 \ln{\left[\frac{a_2(Q^2)}{a_2(M_Z^2)}
\sqrt{\frac{1 + b_1a_2(M_Z^2) + b_2a_2^2(M_Z^2)}
{1 + b_1a_2(Q^2) + b_2a_2(Q^2)}}\right]} \\ \nonumber
&+& \left(b_2-\frac{b_1^2}{2}\right)\times
\Bigl(I(a_s(Q^2))- I(a_s(M_Z^2))\Bigr) = \beta_0 \ln{\left(\frac{Q^2}{M_Z^2}\right)}\,.
\eea
The expression for $I$ looks:
$$
I(a_s(Q^2))=\cases{
\displaystyle{\frac{2}{\sqrt{\Delta}}} \arctan{\displaystyle{\frac{b_1+2b_2a_2(Q^2)}{\sqrt{\Delta}}}} &for $f=3,4,5; \Delta>0$,\cr
\displaystyle{\frac{1}{\sqrt{-\Delta}}}\ln{\left[
\frac{b_1+2b_2a_2(Q^2)-\sqrt{-\Delta}}{b_1+2b_2a_2(Q^2)+\sqrt{-\Delta}}
\right]}&for $f=6;\quad\Delta<0$, \cr
}
$$
where $\Delta=4b_2 - b_1^2$ and $b_i=\beta_i/\beta_0$ are taken from the QCD $\beta$-function:
$$
\beta(\ar) ~=~ -\beta_0 \ar^2 - \beta_1 \ar^3 - \beta_2 \ar^4 +\ldots \,.
$$

\subsection{$Q^2$-dependence of SF moments}

Wilson coefficient function $C(n,\ar(Q^2))$ is expressed in terms of the coefficients $B_j(n)$ (hereafter (j=1,2)), which are exactly known
(see, e.g. \cite{Shaikhatdenov:2009xd})\footnote{For the odd $n$ values, coefficients $B_j(n)$ and $Z_j(n)$ can be obtained
 by using the analytic continuation~\cite{KaKo}.}
\be
C(n, \ar(Q^2)) = 1 
+ \ar(Q^2) B_{1}(n)
+ \ar^2(Q^2) B_{2}(n) + {\cal O}(\ar^3)\,.
\label{1.cf}
\ee

The $Q^2$-evolution of the PDF moments can be calculated within the framework
of perturbative QCD (see e.g.~\cite{Buras}):
\be
\frac{{\bf f}(n,Q^2)}{{\bf f}(n,Q_0^2)}=\left[\frac{\ar(Q^2)}
{\ar(Q^2_0)}\right]^{\frac{\gamma_{0}(n)}{2\beta_0}}
\times \frac{h(n, Q^2)}{h(n, Q^2_0)}
\,, 
\label{3}
\ee
where
\be\label{hnns}
h(n, Q^2)  = 1 + \ar(Q^2) Z_{1}(n) + \ar^2(Q^2) Z_{2}(n)
+ {\cal O}\left(\ar^3\right)\,,
\ee
and
\begin{eqnarray}
Z_{1}(n) &=& \frac{1}{2\beta_0} \biggl[ \gamma_{1}(n) -
\gamma_{0}(n)\, b_1\biggr]\,, \nonumber \\
Z_{2}(n)&=& \frac{1}{4\beta_0}\left[
\gamma_{2}(n)-\gamma_{1}(n)b_1 + \gamma_{0}(n)(b^2_1-b_2) \right]
+  \frac{1}{2} Z^2_{1}(n)
\,
\label{3.21}
\end{eqnarray}
are combinations of the NLO and NNLO anomalous dimensions $\gamma_{1}(n)$ and $\gamma_{2}(n)$.

For large $n$ (this corresponds to large $x$ values), the coefficients $Z_{j}(n)\sim \ln n$ and $B_{j}(n)\sim \ln^{2j}n$.
So, the terms $\sim B_{j}(n)$ may lead to potentially large contributions and, therefore, should be resummed.

\subsection{Factorization $\mu_F$  scale
}

We intend to consider the dependence of the results on the factorization $\mu_F$
scale caused by (see, e.g.,~\cite{Shaikhatdenov:2009xd}) the truncation of a perturbative series when performing the calculus.
The modification is achieved by replacing $\ar(Q^2)$ in Eq.~(\ref{3.a}) with the expressions in which the scale was accounted in the following way:
$\mu^2_F = k_F Q^2$.

Then, Eq.~(\ref{3.a}) takes the form:
\be
M_n(Q^2) = R_{NS}(f) \times \hat{C}(n, \ar(k_F Q^2))
\times {\bf f}(n,k_F Q^2)\,. \nonumber
\ee

The function $\hat{C}$ is to be obtained from $C$ by modifying the RHS of Eq.~(\ref{1.cf}) as follows:
\bea
\ar(Q^2) &\to& \ar(k_F Q^2)\,,~~
B_{1}(n) \to B_{1}(n) + \frac{1}{2}\gamma_{0}(n) \ln{k_F}\,, \nonumber \\
B_{2}(n) &\to& B_{2}(n) + \frac{1}{2}\gamma_{1}(n) \ln{k_F}
+\left(\frac{1}{2}\gamma_{0} + \beta_0\right)B_{1}\ln{k_F}
+\frac{1}{8}\gamma_{0}\left(\gamma_{0} + 2\beta_0\right)\ln^2{k_F}\,.
\label{coeffun}
\eea

Taking a special form for the coefficient $k_F$, we can decrease contributions coming from the terms $\sim B_{j}(n)$.


\section{DIS scheme}

Let us consider the case of the so-called DIS-scheme \cite{DIS} (it was also called the $\Lambda_n$-scheme \cite{Lambda_n}),
where NLO corrections to the Wilson coefficients are completely compensated by the factorization scale variation.

\subsection{NLO}

In this order
\be
a_s(Q^2) \to a_s(k_{\rm DIS}(n)Q^2)\equiv a_n^{\rm DIS}(Q^2),~~
k_{\rm DIS }(n)=exp\left(\frac{-2B_{1}(n)}{\gamma_0(n)}\right) =exp\left(\frac{-r^{\rm DIS }_{1}(n)}{\beta_0}\right) \, ,
\label{kDIS.NLO}
\ee
where
\be
r^{\rm DIS }_{1}(n)=\frac{2B_{1}(n)\beta_0}{\gamma_0}
~~~
\mbox{and}~~~
 B_1(n) \to B^{\rm DIS}_{1} =0,
\label{oBDI.NLO}
\ee
i.e. $\hat{C}(n, a_n^{\rm DIS}(Q^2))=1+ {\cal O}((a_n^{\rm DIS})^2)$.

The NLO coupland $a^{\rm DIS}_n(Q^2)$ obeys the following equation
\bea \label{1.coA.DIS}
&&\frac{1}{a^{\rm DIS}_{n}(Q^2)} - \frac{1}{a_{1}(M_Z^2)} +
b_1 \ln{\left[\frac{a^{\rm DIS}_{n}(Q^2)}{a_{1}(M_Z^2)}
\frac{(1 + b_1a_{1}(M_Z^2))}
{(1 + b_1a^{\rm DIS}_{n}(Q^2))}\right]}= \beta_0 \ln{\left(\frac{k_{\rm DIS}(n)Q^2}{M_Z^2}\right)}\nonumber \\
&&\hspace{1cm}  = \beta_0 \ln{\left(\frac{Q^2}{M_Z^2}\right)}-r^{\rm DIS}_{1}(n)\,.
\eea

\subsection{NNLO}

In this case we have Eqs. (\ref{kDIS.NLO}) and (\ref{oBDI.NLO}) and additionally
\be
B_2(n) \to B^{\rm DIS}_2(n)= B_2(n)-\left(\frac{1}{2}+\frac{\beta_0}{\gamma_0(n)}\right)\, B^2_1(n)-
\frac{\gamma_1(n)}{\gamma_0(n)}\, B_1(n)\, ,
\label{oBDI.NNLO}
\ee
that leads to the cancellation of
the larger terms $\sim \ln^4(n)$
in $B^{\rm DIS}_2(n)$.

The NNLO coupland $a^{\rm DIS}_n(Q^2)$ obeys the following equation
\bea \label{1.coB.DIS}
\frac{1}{a_n(Q^2)} - \frac{1}{a_2(M_Z^2)} &+&
b_1 \ln{\left[\frac{a_n(Q^2)}{a_2(M_Z^2)}
\sqrt{\frac{1 + b_1a_2(M_Z^2) + b_2a_2^2(M_Z^2)}
{1 + b_1a_n(Q^2) + b_2a_n(Q^2)}}\right]} \\ \nonumber
&& \hspace{-2cm} +\left(b_2-\frac{b_1^2}{2}\right)\times
\Bigl(I(a_n(Q^2))-I(a_s(M_Z^2))\Bigr) = \beta_0 \ln{\left(\frac{Q^2}{M_Z^2}\right)}-r^{\rm DIS}_{1}(n)\,.
 \eea

\section{ A fit method }

The most popular approach (see e.g.~\cite{NNLOfits}) to carrying out QCD analyses over DIS data is associated
with Dokshitzer-Gribov-Lipatov-Altarelli-Parisi (DGLAP) integro-differential equations~\cite{DGLAP}.
It is a brute force method and allows one to analyze the data directly.

However, as seen from our previous efforts we advocate another approach. 
The analysis is carried out with the moments of SF $F_2(x,Q^2)$ defined in Eq. (\ref{Moments}).
Then, for each $Q^2$, SF
$F_2(x,Q^2)$ is recovered using the Jacobi polynomial decomposition method \cite{Barker,Kri}:
\be
F_{2}(x,Q^2)=x^a(1-x)^b\sum_{n=0}^{N_{max}}\Theta_n ^{a,b}(x)\sum_{j=0}^{n}c_j^{(n)}(\alpha ,\beta )
M_{j+2} (Q^2)\,,
\label{2.1}
\ee
where $\Theta_n^{a,b}$ are the Jacobi polynomials, $a,b$ are the parameters to be fit.
As usual, the compliance condition is the requirement to minimize the error in restoring the structure functions.

The program MINUIT \cite{MINUIT} is used to minimize the variable
\be
\chi^2_{SF} = \biggl|\frac{F_2^{exp} - F_2^{th}}{\Delta F_2^{exp}}\biggr|^2\,.
\label{chi2}
\ee

\section{Results}

We use free data normalizations for various experiments. As a reference set, the most stable hydrogen BCDMS data are used at the value of
the initial beam energy $E_0=200$ GeV.
Contrary to previous analyses \cite{Shaikhatdenov:2009xd,Kotikov:2015zda}, the cut $Q^2\geq 2$GeV$^2$ is used throughout, since for
smaller $Q^2$ values the equations (\ref{1.coA.DIS}) and (\ref{1.coB.DIS}) have no real solutions.

The starting point of $Q^2$-evolution is taken at
$Q^2_0$ = 90 GeV$^2$. This particular value of $Q^2_0$ is close to the average values of $Q^2$ covering the corresponding data.
Based on previous investigations (see Ref. \cite{Kri}), the maximum number of moments to be accounted for is $N_ {max} = $8, and
the cut $0.25 \leq x \leq 0.8$ is applied on the data.

We work within the framework of the variable-flavor-number scheme (VFNS) (see \cite{Shaikhatdenov:2009xd}). 
To make it more clear the effect of changing the sign for twist-four corrections, the fixed-flavor-number scheme with $n_f=4$ is also used.

\vspace{0.5cm}
       {\bf Table 1.}  Parameter values of the twist-four term in different cases obtained in the analysis of data (314 points: $Q^2\geq 2$ GeV$^2$)
       carried out within VFNS
       (FFNS). 
\footnotesize
       \begin{center}
\begin{tabular}{|l|c|c|c|c|c|}
\hline
&NLO &NLO &NNLO &NNLO  \\
$x$  & $\overline{MS}$ scheme & DIS scheme           &  $\overline{MS}$ scheme & DIS scheme \\
 &$\chi^2=246 (259)$ &  $\chi^2=238 (251)$ &$\chi^2=241 (254)$  &  $\chi^2=242(249)$    \\
&$\alpha_s(M_Z^2)= 0.1195$ &  $\alpha_s(M_Z^2)= 0.1177$ &$\alpha_s(M_Z^2)= 0.1177$  &  $\alpha_s(M_Z^2)= 0.1178$    \\
&(0.1192) &(0.1179)&(0.1170) &(0.1171)\\
\hline \hline
0.275 & -0.25$\pm$0.02 (-0.26$\pm$0.03) & -0.18$\pm$0.01 (-0.17$\pm$0.02) & -0.19$\pm$0.02 (-0.20$\pm$0.02) & -0.14$\pm$0.01 (-0.17$\pm$0.01)  \\
0.35 & -0.24$\pm$0.02 (-0.25$\pm$0.02) &  -0.11$\pm$0.01 (-0.13$\pm$0.01)  & -0.19$\pm$0.03 (-0.19$\pm$0.02) &  -0.13$\pm$0.02 (-0.15$\pm$0.01)  \\
0.45 & -0.19$\pm$0.02 (-0.19$\pm$0.02) &  -0.04$\pm$0.04 (-0.09$\pm$0.01) & -0.17$\pm$0.03 (-0.16$\pm$0.01) &  -0.11$\pm$0.09 (-0.10$\pm$0.02)  \\
0.55 & -0.12$\pm$0.03 (-0.10$\pm$0.03)&  -0.11$\pm$0.01 (-0.09$\pm$0.04) & -0.17$\pm$0.05 (-0.14$\pm$0.03) &  -0.12$\pm$0.03 (-0.08$\pm$0.04)  \\
0.65 & 0.05$\pm$0.08 (0.12$\pm$0.08) &  -0.17$\pm$0.04 (-0.09$\pm$0.05) &  -0.14$\pm$0.14 (-0.05$\pm$0.06) &  -0.22$\pm$0.05 (-0.10$\pm$0.05) \\
0.75 & 0.34$\pm$0.12 (0.48$\pm$0.12) & -0.57$\pm$0.08 (-0.44$\pm$0.18) & -0.11$\pm$0.19 (0.06$\pm$0.10) & -0.59$\pm$0.08 (-0.32$\pm$0.12)  \\
\hline
\end{tabular}
\end{center}
\vspace{0.5cm}
\normalsize

As is seen from Table 1 the central values of $\alpha_s(M_Z^2)$ are fairly the same given total experimental and theoretical errors
(see \cite{Shaikhatdenov:2009xd,Kotikov:2015zda}):
\be
\pm 0.0022~~~\mbox{(total exp. error)},~~~
\cases{
  \displaystyle{+0.0028}\cr
\displaystyle{-0.0016}\cr
}
~~~\mbox{(theor. error)}\,.
\label{Errors}
\ee
We plan to study the errors in more detail and present them in an upcoming publication.

From Table 1, it can also be seen that upon resumming at large $x$ values (i.e. in the DIS scheme), the twist-four corrections
become large and negative
in this $x$ region.
Moreover, it seems that they rise as $1/(1-x)$ at large $x$ but this observation needs additional investigations.

Such a behavior is completely contrary to the analyses \cite{Shaikhatdenov:2009xd,Kotikov:2015zda,Blumlein:2008kz,Alekhin:2000ch} performed in
  $\overline{MS}$ scheme, where twist-four corrections are mostly positive at large $x$ and rise as $1/(1-x)$. Note that this rise is usually 
less pronounce
  in higher orders (see \cite{Shaikhatdenov:2009xd,Kotikov:2015zda,Blumlein:2008kz}) and sometimes is even absent at NNLO level (see Table 1).

The negative values of powerlike corrections at large $x$ obtained in DIS scheme leads to the following phenomenon:
(part of) powerlike
terms can be absorbed into the difference between usual strong coupling and QCD analytical one
\cite{SoShi} just the same way as it was done at low $x$ values
(see Refs. \cite{CIKK09,Kotikov:2012sm}) in the framework of the so-called double asymptotic scaling approach \cite{Q2evo}.
    Of course, such a phenomenon was absent in the case of $\overline{MS}$ scheme, where using \cite{Kotikov:2010bm}
       the QCD analytical coupling simply increases the magnitude of twist-four corrections.

In previous papers (see \cite{Vovk,PKK}), where resumming at large values of $x$ was performed within the framework 
of the Grunberg approach \cite{Grunberg},  only a decrease in the twist-four contribution was seen, 
since the relevant terms were not studied in detail. Therefore, it looks promising if the Grunberg approach will be used
in the analysis similar to the present one and thus promote the study in some detail of the twist-four correction values.

\section{Summary}

We have performed fits of experimental data of BCDMS, SLAC and NMC collaborations for DIS SF $F_2(x,Q^2)$ by resumming 
large logarithms at large $x$ values into the corresponding coefficient function within the DIS scheme.

It is seen that the resummation does not change the values of the strong coupling constant $\alpha_s(M_Z^2)$, 
though the values of the twist-four corrections become large and negative contrary to the results obtained in the
  $\overline{MS}$ scheme analyses.
We plan to study this phenomenon using another scheme of resumming large logarithms at large $x$ values,
such as the Grunberg approach \cite{Grunberg}.\\

A.V.K. thanks the Organizing Committee of the International Conference ``alphas-2022: Workshop on precision measurements of the QCD coupling constant''
for invitation.

\end{document}